\documentclass[twocolumn, journal]{IEEEtran}
\IEEEoverridecommandlockouts
\usepackage{color}
\usepackage{graphicx}
\usepackage{epstopdf}
\usepackage{amsmath}
\usepackage{amssymb}
\usepackage{algorithm}
\usepackage{algorithmic}
\usepackage{amsmath}
\usepackage{multirow}
\usepackage{booktabs}
\usepackage{array}
\usepackage{amsthm}
\usepackage{stfloats}
\usepackage{caption}
\usepackage{subfigure}
\usepackage{bm}
\usepackage{setspace}
\usepackage{diagbox}
\usepackage{flushend}
\usepackage{enumitem}
\usepackage[table]{xcolor}
\definecolor{myblue}{RGB}{213, 224, 244}

\newcommand{\be}{\begin{equation}}
\newcommand{\ee}{\end{equation}}
\newcommand{\bea}{\begin{eqnarray}}
\newcommand{\eea}{\end{eqnarray}}
\newcommand{\ba}{\begin{array}}
\newcommand{\ea}{\end{array}}

\allowdisplaybreaks[4]

\def\BibTeX{{\rm B\kern-.05em{\sc i\kern-.025em b}\kern-.08em
    T\kern-.1667em\lower.7ex\hbox{E}\kern-.125emX}}
\begin{document}

\title{Reconfigurable Antenna Arrays: Bridging Electromagnetics and Signal Processing
%\title{Integrating Electromagnetic and Signal Processing Domains: Reconfigurable Antenna Array in 6G Networks
%\thanks{Mengzhen Liu, Ming Li, and Qian Liu are with Dalian University of Technology, China. Rang Liu and Lee Swindlehurst are with University of California, USA.}
}

\author{Mengzhen Liu,~\IEEEmembership{Graduate Student Member,~IEEE,}
        Ming Li,~\IEEEmembership{Senior Member,~IEEE,}
        Rang Liu,~\IEEEmembership{Member,~IEEE,}\\
        Qian Liu,~\IEEEmembership{Member,~IEEE,}
        and A. Lee Swindlehurst,~\IEEEmembership{Life Fellow,~IEEE}\\
\vspace{-0.0 cm}
}

\maketitle
\pagestyle{empty}  % no page number for the second and the later pages
\thispagestyle{empty} % no page number for the first page

\begin{abstract}
Reconfigurable antennas (RAs), capable of dynamically adapting their radiation patterns, polarization states, and operating frequencies, have emerged as a promising technology to meet the stringent performance requirements of sixth-generation (6G) wireless networks. This article systematically introduces essential hardware implementations of RAs and investigates advanced array architectures, such as fully-digital and tri-hybrid designs, emphasizing their capability to synergistically integrate electromagnetic (EM) reconfigurability with analog and digital signal processing. By facilitating coordinated beamforming across the EM and signal processing domains, RA arrays offer unprecedented flexibility and adaptability compared to conventional static antenna systems. Representative applications empowered by RA arrays, including integrated sensing and communication (ISAC), physical layer security (PLS), and near-field communications, are highlighted. A case study illustrates the effectiveness of RA arrays in optimizing beam steering, improving link robustness, and alleviating system power consumption. Finally, several open challenges and future research directions are outlined, emphasizing the need for advancements in theoretical modeling, hardware reliability, channel estimation techniques, intelligent optimization methods, and innovative network architectures, to fully realize the transformative impact of RAs in future 6G wireless networks.
\end{abstract}


\vspace{-0.3cm}
\section{Introduction}
Sixth-generation (6G) wireless systems are expected to provide extremely high data rates, ultra-reliable connectivity, negligible latency, and ubiquitous coverage, thereby transforming exising communication paradigms. In previous wireless generations, these requirements were primarily addressed by scaling up antenna array sizes, as exemplified by the development of massive multiple-input multiple-output (MIMO) systems. However, continuously increasing array size introduces significant challenges, including markedly higher hardware complexity, elevated power consumption, and increased deployment difficulty. Consequently, exploring new degrees of freedom (DoFs) is essential to overcome these performance bottlenecks and fully unlock the potential of next-generation networks.
Recently, fluid antennas (FAs) have emerged as a promising means of introducing additional DoFs by enabling dynamic antenna repositioning, but their real-time adaptability, fabrication complexity, and mechanical costs hinder large-scale practical deployment.

Alternatively, integrating reconfigurable antennas (RAs) into MIMO transceivers offers another compelling approach to creating new DoFs, significantly expanding traditional antenna design paradigms \cite{J. Costantine reconfigurable Proc. IEEE}. Unlike conventional static antennas, RAs can dynamically adjust their fundamental electromagnetic (EM) properties, including radiation patterns, polarization states, and operating frequencies, in response to changing channel conditions and diverse operational requirements. This ability allows RAs to flexibly perform multiple functions, significantly enhancing system agility, adaptability, and responsiveness.

Given their inherent advantages in EM reconfigurability, high energy efficiency, and compact form factor, RA arrays present a particularly attractive solution for future wireless systems. By replacing a large number of conventional static antenna elements with a smaller number of RAs, one can achieve comparable system performance with reduced array size, power consumption, and hardware complexity. RA arrays offer unique capabilities for per-element EM reconfigurability, establishing a tightly integrated interface between EM-domain control and radio frequency (RF)/baseband (BB)-domain signal processing (SP). By jointly exploiting tunable EM properties and sophisticated SP techniques, cross-domain optimization enables significant performance enhancements that surpass those achievable through independent domain-specific designs. This integrated approach facilitates the realization of more precise beamforming, improved spatial multiplexing, and enhanced interference suppression, collectively advancing the overall performance and adaptability of wireless systems beyond the inherent limitations of conventional architectures.

Widespread deployment of RA arrays that bridge the EM and SP domains requires that several critical theoretical and practical challenges be addressed. To describe these challenges, in this paper we introduce the fundamental operating principles of RAs and their adaptability, and describe  their core radiation characteristics. Building on these foundational concepts, we investigate the integration of RAs into antenna array architectures, highlighting the methodologies and advantages of coordinated designs across the two domains. To illustrate the practical implications of this integrated approach, we explore representative applications and present a detailed case study that highlights the tangible performance enhancements that result from RA arrays and cross-domain integration. We also outline several open research challenges and future directions, underscoring the transformative role that RA arrays may play in shaping future wireless communication systems.

\vspace{-0.2cm}
\section{Principles and Characteristics of RAs}\label{Hardware}

\begin{figure*}[!t]
   \centering
    \vspace{-0.1cm}
    \subfigure[Pattern reconfiguration.]{\includegraphics[width= 1.8 in]{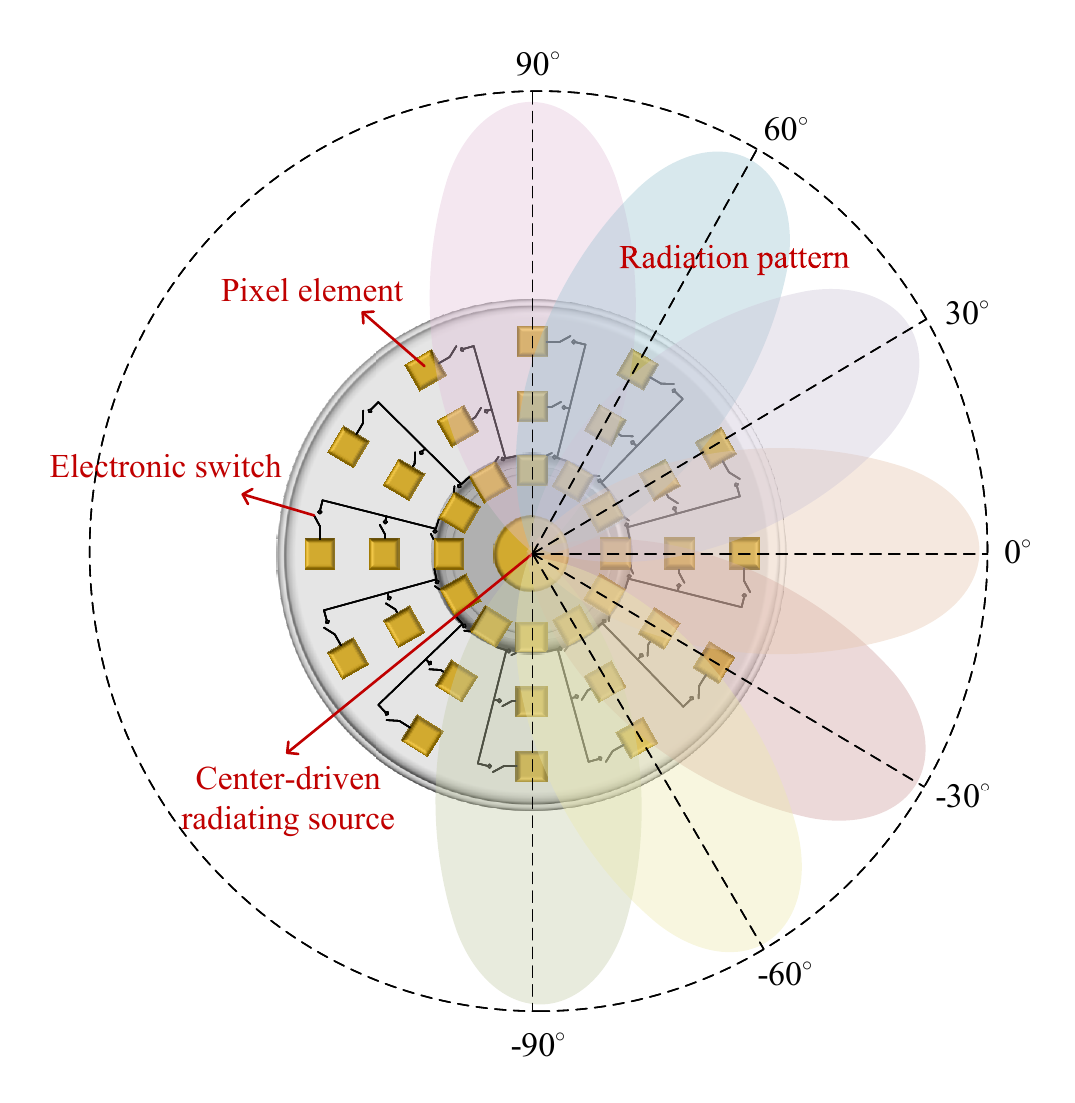}}
    \hspace{0.3 cm}
    \subfigure[Polarization reconfiguration.]{\includegraphics[width= 1.8 in]{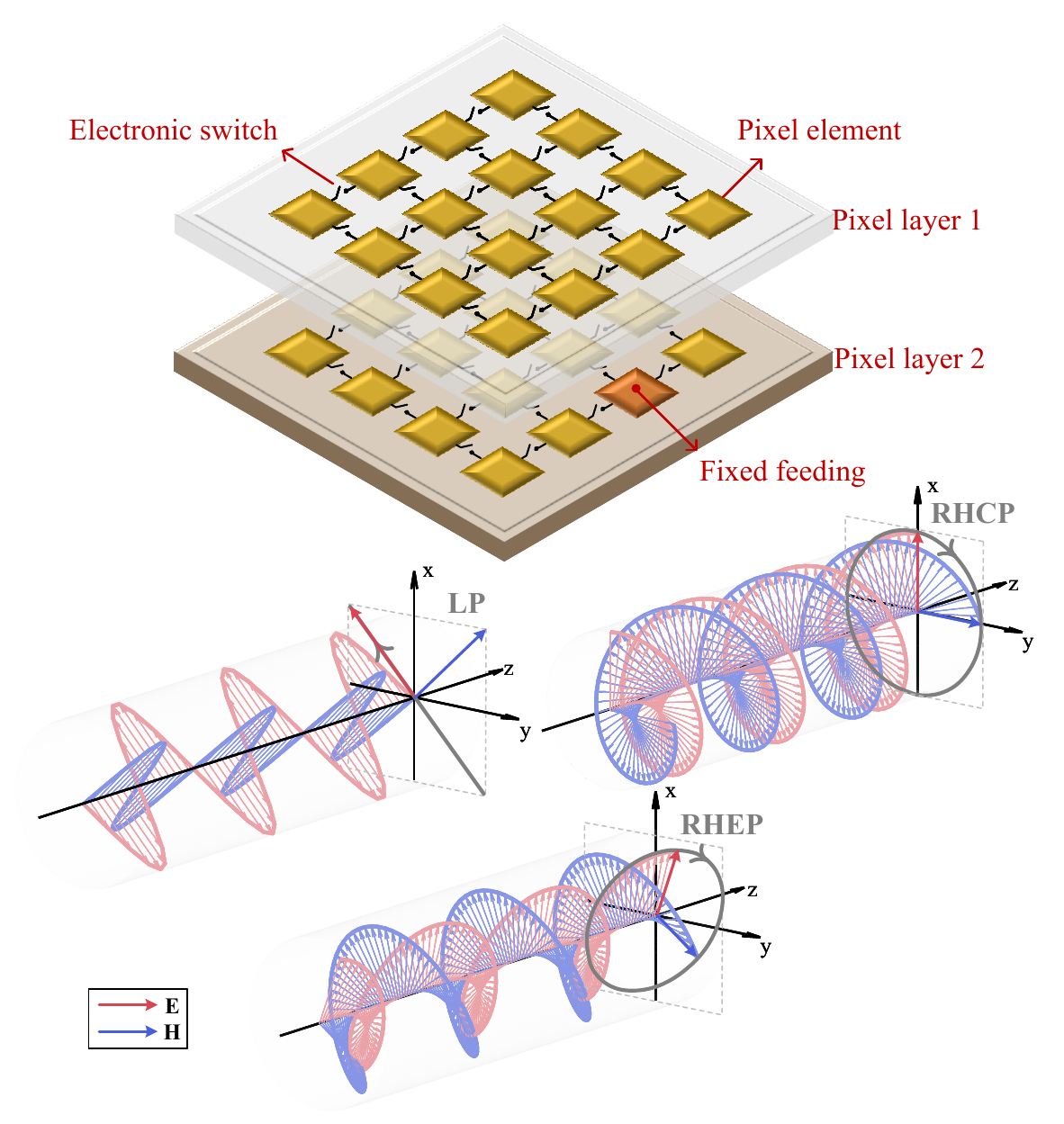}}
    \subfigure[Frequency reconfiguration.]{\includegraphics[width= 2.0 in]{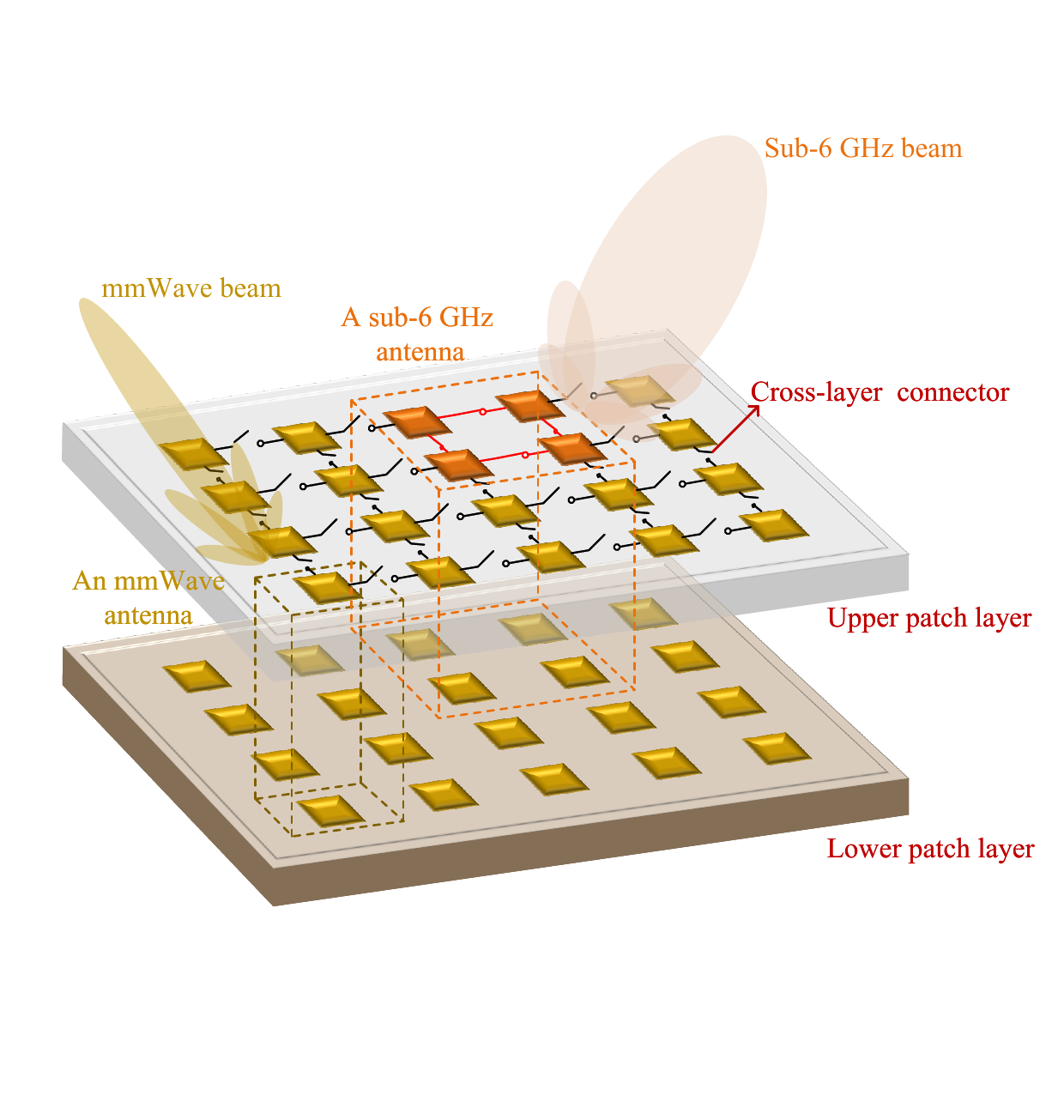}}
    \caption{RA hardware architectures.}
     \vspace{-0.4 cm}
    \label{fig:hardware}
\end{figure*}

%\subsection{Principles of Reconfigurability}
%RA represents a revolutionary breakthrough in modern antenna technology, distinguished by its unique capability to dynamically modify EM characteristics, primarily including radiation pattern, polarization state, and operating frequency. In the following, we will systematically elaborate on the operating principles of RAs from both the perspective of physical structure and theoretical model.

Antenna reconfigurability is typically realized using electrical, optical, mechanical, or materials-based methods. Among these, electrical approaches are the most prevalent due to their compact size, fast response, and compatibility with modern RF systems, such as electronically steerable parasitic array radiator (ESPAR), varactor-tuned antenna, and pixel-based antenna \cite{hardware classification}. This section focuses on the underlying principles of electrical RAs, which are broadly applicable to other reconfiguration methods as well.

From an EM perspective, an RA is generally regarded as a structure comprising a primary radiating source and multiple parasitic elements (e.g., pixels) that are interconnected via controllable electronic switches, as shown in Fig.~\ref{fig:hardware}. By selectively toggling the switch states, the internal topology of the RA is dynamically altered, thereby influencing the EM coupling among the constituent elements. Moreover, the current distribution across the structure will be modified, resulting in changes in the antenna’s radiated EM field and its associated properties. The following subsections detail three fundamental reconfiguration modes.

\vspace{-0.4cm}
\subsection{Pattern Reconfigurability}
Radiation patterns define the spatial distribution of EM energy and directly impact directionality, beamwidth, and gain. The patterns are typically categorized as omnidirectional, directional, or reconfigurable. While omnidirectional antennas offer broad coverage with low gain, directional antennas concentrate energy into narrow beams. Pattern-RA, as shown in Fig.~\ref{fig:hardware}(a), can balance coverage and radiation intensity by adaptively reshaping beams to match changing environments and operational demands. This adaptability enables dynamic pattern steering toward desired directions while suppressing interference from undesired angles \cite{R. Murch 2022 pixel fig_reference_round}.

\vspace{-0.4cm}
\subsection{Polarization Reconfigurability}
Polarization describes the orientation and trajectory of the electric field vector, and is a fundamental antenna attribute. It is usually classified as linear (horizontal or vertical), circular (right-hand or left-hand), or elliptical, as depicted in Fig.~\ref{fig:hardware}(b) \cite{H. Li 2024 polarization}. Polarization inherently serves as a spatial filtering mechanism: Effective reception occurs when the signal’s polarization aligns with that of the antenna. Traditional dual-polarized antennas achieve polarization diversity via two orthogonally polarized radiators, each requiring a separate RF chain. In contrast, a polarization-RA utilizes a single radiator capable of dynamically generating multiple polarization states, thus achieving diversity benefits without the hardware complexity or energy cost of additional RF chains.

\vspace{-0.4cm}
\subsection{Frequency Reconfigurability}
Frequency reconfigurability enables efficient operation over multiple bands, supporting agile spectrum access. Conventional designs include narrowband, wideband, and discrete multiband antennas, each with trade-offs in gain, selectivity, and hardware complexity. As depicted in Fig.~\ref{fig:hardware}(c), a frequency-RA can adaptively switch among bands using a unified structure, thereby reducing component redundancy and improving spectral efficiency \cite{R. Murch 2024 DUAL}.

\vspace{-0.2cm}
\section{RA Array Architectures and Multi-domain Designs}\label{Architecture}
Owing to their flexible EM-domain reconfigurability, low power consumption, and compact form factor, RAs are ideally suited for integration into antenna arrays. RA arrays can achieve comparable or superior performance to conventional designs while significantly reducing array size, system complexity, and energy consumption. Motivated by these advantages, this section investigates the paradigm shift in array design driven by RA technology and introduces a coordinated design framework to effectively bridge EM and SP.

\begin{figure*}[!t]
   \centering
    \hspace{-0.4cm}
    \subfigure[Fully-digital.]{\includegraphics[height= 1.75 in]{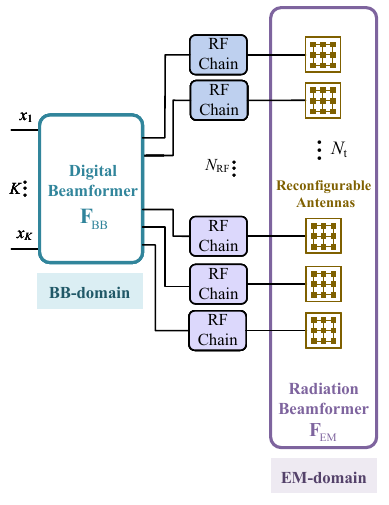}}
    \hspace{0.3cm}
    \subfigure[Fully-connected tri-hybrid.]{\includegraphics[height= 1.75 in]{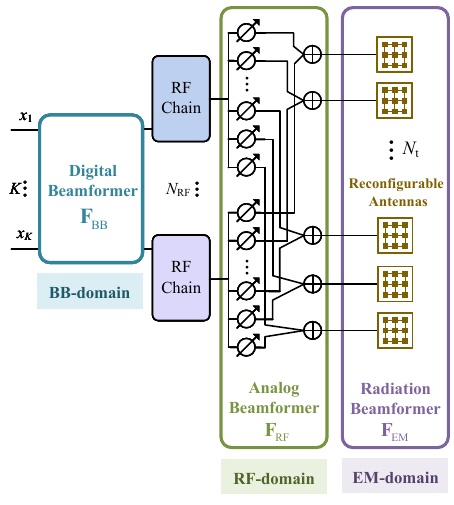}}
    \hspace{0.3cm}
    \subfigure[Sub-connected tri-hybrid.]{\includegraphics[height= 1.75 in]{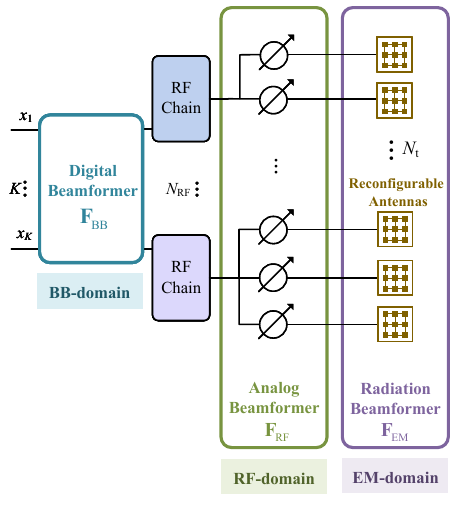}}
    \hspace{0.3cm}
    \subfigure[Dynamic-connected tri-hybrid.]{\includegraphics[height= 1.75 in]{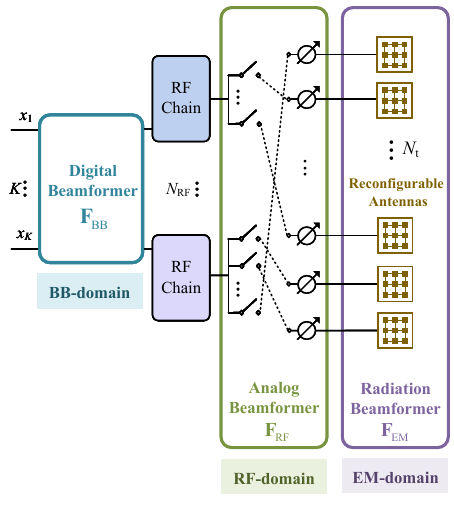}}
    \caption{The representative RA array architectures.}
     \vspace{-0.4 cm}
    \label{fig:architecture}
\end{figure*}

\vspace{-0.3cm}
\subsection{RA Array Architectures}
Fig.~\ref{fig:architecture}(a) illustrates a fully-digital beamforming architecture integrated with an RA array. In this design, each RA is connected to a dedicated RF chain, enabling a two-stage design: (\textit{i}) an EM-domain beamformer to adaptively control radiation characteristics, and (\textit{ii}) a BB-domain digital beamformer to achieve precise beam alignment. Such RA array architectures significantly extend the capabilities of traditional fully-digital beamforming architectures, delivering enhanced flexibility through joint optimization of EM radiation and digital beamforming.

Despite these performance gains, the growing number of antenna elements and the increasing operational frequency ranges pose significant implementation challenges for next-generation wireless systems. To address these concerns, hybrid beamforming architectures have emerged as a power- and cost-efficient alternative to fully-digital architectures, reducing the number of required RF chains while maintaining high performance. Recent works have further advanced this approach by integrating RA technology into hybrid architectures, giving rise to so-called ``\textbf{tri-hybrid}'' beamforming schemes \cite{R. W. Heath 2025 tri-hybrid}. These architectures leverage the reconfigurability of RAs to enhance beamforming adaptability, while simultaneously benefiting from the hardware efficiency and reduced power consumption inherent to hybrid designs.

As illustrated in Fig.~\ref{fig:architecture}(b)-(d), representative tri-hybrid architectures typically adopt a three-layer structure: (\textit{i}) an EM-domain beamformer to tailor the antenna's EM characteristics, (\textit{ii}) an RF-domain analog beamformer to enhance array gain, and (\textit{iii}) a BB-domain digital beamformer to mitigate multi-user interference. Depending on the connectivity between the RF chains and phase shifters (PSs), tri-hybrid architectures can be further classified as fully-connected, sub-connected, or dynamic-connected implementations.

\vspace{-0.3cm}
\subsection{Multi-domain Coordinated Design}
Beyond architectural innovations, RA arrays call for a coordinated multi-domain design framework that facilitates joint optimization across the EM, RF, and BB domains. This co-design is inherently coupled: signal processing in the RF/BB domains can guide EM mode selection, while feedback from EM characteristics can refine subsequent signal processing. This iterative and synergistic interaction progressively aligns all domains toward a globally optimal configuration.

Practical demonstrations of such coordinated frameworks have been presented in \cite{C.-B. Chae 2025 tri-hybrid},  where EM-domain unit realized through the dynamic metasurface antenna (DMA), interface seamlessly with RF and BB modules to enable efficient cross-domain beamforming and achieve higher energy efficiency than traditional architectures. Additionally, a tri-hybrid design in \cite{M. Liu 2025 tri-hybrid} employs a novel tri-timescale optimization strategy that matches the temporal dynamics of each domain. These examples highlight the practical feasibility and performance benefits of coordinated multi-domain designs, showcasing the transformative potential of RA arrays in future wireless systems.

\vspace{-0.2cm}
\section{Applications Empowered by RA Arrays}\label{Application}
Leveraging the flexible EM reconfigurability enabled by RA arrays, recent research has demonstrated their significant potential in enhancing wireless system performance. In particular, joint optimization of the radiation pattern and beamforming of an RA array has been shown to improve channel capacity \cite{C.-B. Chae 2025 tri-hybrid}, \cite{R. Murch 2023 ESPAR}, \cite{K.K.Wong R. Murch 2024 Antenna coding pixel}. Additionally, antenna polarization reconfigurability has demonstrated enhanced polarization diversity and improved link robustness \cite{Z. Zhou 2025 ISAC}. Beyond communication-centric gains, this section outlines several other representative applications enabled by RA arrays, as summarized in Table I, followed by a detailed case study.

\begin{table*}[htb]
\begin{small}
\caption{Summary of research on RA arrays.}
\vspace{-0.3cm}
\end{small}
\small
\begin{center}
\footnotesize
\begin{tabular}{>{\columncolor{myblue}}>{\centering\arraybackslash}m{1.6cm} >{\columncolor{gray!15}}>{\centering\arraybackslash}m{0.6cm}
>{\columncolor{myblue}}>{\centering\arraybackslash}m{1.9cm}
>{\columncolor{gray!15}}>{\centering\arraybackslash}m{2.1cm}
>{\columncolor{myblue}}>{\centering\arraybackslash}m{4.2cm}
>{\columncolor{gray!15}}>{\centering\arraybackslash}m{4.5cm}}
\textbf{Applications} & \textbf{Ref.} & \textbf{Hardware} & \textbf{Reconfigurable Characteristics}  & \textbf{EM Domain Optimization Variables} & \textbf{Signal Processing Domain Optimization Variables} \\
\hline
& \cite{C.-B. Chae 2025 tri-hybrid} & DMA & Pattern & Phase shift and amplitude of each DMA element & Hybrid beamforming \\
\cline{2-6}
& \cite{R. Murch 2023 ESPAR}& ESPAR & Pattern & Reactive loads & Analog beamforming \\
\cline{2-6}
 &  \cite{K.K.Wong R. Murch 2024 Antenna coding pixel} & Pixel & Pattern &  Antenna coding &Transmit signal covariance  \\
\cline{2-6}
\multirow{-5}*{Commun.}  &  \cite{Z. Zhou 2025 ISAC}  & PS-based RAs &  Polarization & Transmit \& receive phase shifts & Transmit signal covariance\\
\hline
& \cite{R. Liu security RA } & / & Polarization &Transmit \& receive polarization & Waveforms \\
\cline{2-6}
\multirow{-2}*{ISAC} & \cite{K. Chen 2025 DUAL}  & Pixel & Frequency & MmWave antennas combination & MmWave hybrid beamforming \& sub-6G fully-digital beamforming\\
\hline
Security & \cite{K. Huang 2023 Security} & DMA & Pattern & Phase shift and amplitude of each DMA element & Digital beamforming
\end{tabular}
\end{center}\vspace{-0.5cm}
\end{table*}

\vspace{-0.4cm}
\subsection{ISAC}
Integrated sensing and communication (ISAC) aims to unify sensing and communication functionalities within a shared hardware and spectral framework. This integration, however, inherently introduces performance trade-offs where enhancements in sensing accuracy may compromise communication throughput and vice versa. RA arrays offer a promising avenue to mitigate these trade-offs by introducing additional EM-domain DoFs that can be jointly optimized with SP algorithms to dynamically manage and balance ISAC performance.

Among various RA features, polarization and frequency reconfigurability are particularly advantageous for ISAC, as illustrated in Fig.~\ref{fig:applications}(a). Polarization reconfigurability enables sensing systems to exploit polarization-dependent scattering characteristics, thereby improving target detection and classification, especially in cluttered or multipath-rich environments. On the communication side, polarization diversity enhances link reliability and spectral efficiency. For instance, \cite{R. Liu security RA } proposes an integrated polarimetric sensing and communication (IPSAC) system, where polarization states and waveforms are co-optimized to improve sensing accuracy while satisfying communication quality-of-service (QoS) constraints.

Frequency reconfigurability further enhances ISAC system performance by enabling dual-band operation, thus accessing a broader spectral range and leveraging complementary propagation characteristics across sub-6 GHz and millimeter-wave (mmWave) bands. As demonstrated in \cite{K. Chen 2025 DUAL}, a dual-band RA array jointly operating in mmWave and sub-6 GHz frequencies facilitates high data-rate transmission and wide-area coverage for communication, while simultaneously supporting high-resolution and large-scale sensing. These capabilities underscore the unique benefits of RA arrays in realizing flexible, high-performance ISAC systems.

\subsection{Physical Layer Security}
Physical layer security (PLS) aims to safeguard wireless
communications by exploiting the inherent characteristics of the physical propagation channel. However, conventional PLS approaches are often limited in highly dynamic or correlated environments. RA arrays provide a promising solution by dynamically tuning key EM characteristics, particularly radiation patterns and polarization states, in order to proactively manipulate the wireless propagation environment and degrade eavesdropper reception while enhancing legitimate link security. For instance, \cite{K. Huang 2023 Security} proposes an anti-jamming framework based on heterogeneous dynamically reconfigurable metasurface arrays in which EM radiation patterns and SP weights are jointly optimized to improve robustness against jamming. Moreover, as shown in Fig.~\ref{fig:applications}(b), polarization reconfigurability enables selective transmission and reception of desired signal polarizations while suppressing the polarization states of eavesdroppers.

%Physical layer security (PLS) aims to safeguard wireless communications by exploiting the inherent characteristics of the physical channel, offering a complementary approach to traditional cryptographic methods. However, conventional PLS techniques often struggle under adverse conditions, such as hostile jamming, rapidly changing environments, and highly correlated channels. RA technology offers a promising solution to these challenges by dynamically tuning key EM characteristics, particularly radiation patterns and polarization, thereby proactively manipulating the wireless propagation environment. This capability allows the system to degrade signal quality at potential eavesdroppers while enhancing the reliability and secrecy of legitimate links.

%For instance, \cite{K. Huang 2023 Security} proposes a robust anti-jamming communication framework based on dynamically pattern-reconfigurable heterogeneous DMA arrays. By jointly optimizing the EM-domain radiation patterns and signal-processing-domain array weights, this design substantially enhances wireless link secrecy and robustness against fluctuating jamming attacks. In addition, as shown in Fig.~\ref{fig:applications}(b), polarization reconfiguration via RAs enables selective reception of signals in desired polarization states, effectively suppressing those polarization states accessible to potential eavesdroppers.

\vspace{-0.4cm}
\subsection{Near-field}
\begin{figure*}[!t]
   \centering
    \subfigure[ISAC.]{\includegraphics[width= 1.7 in]{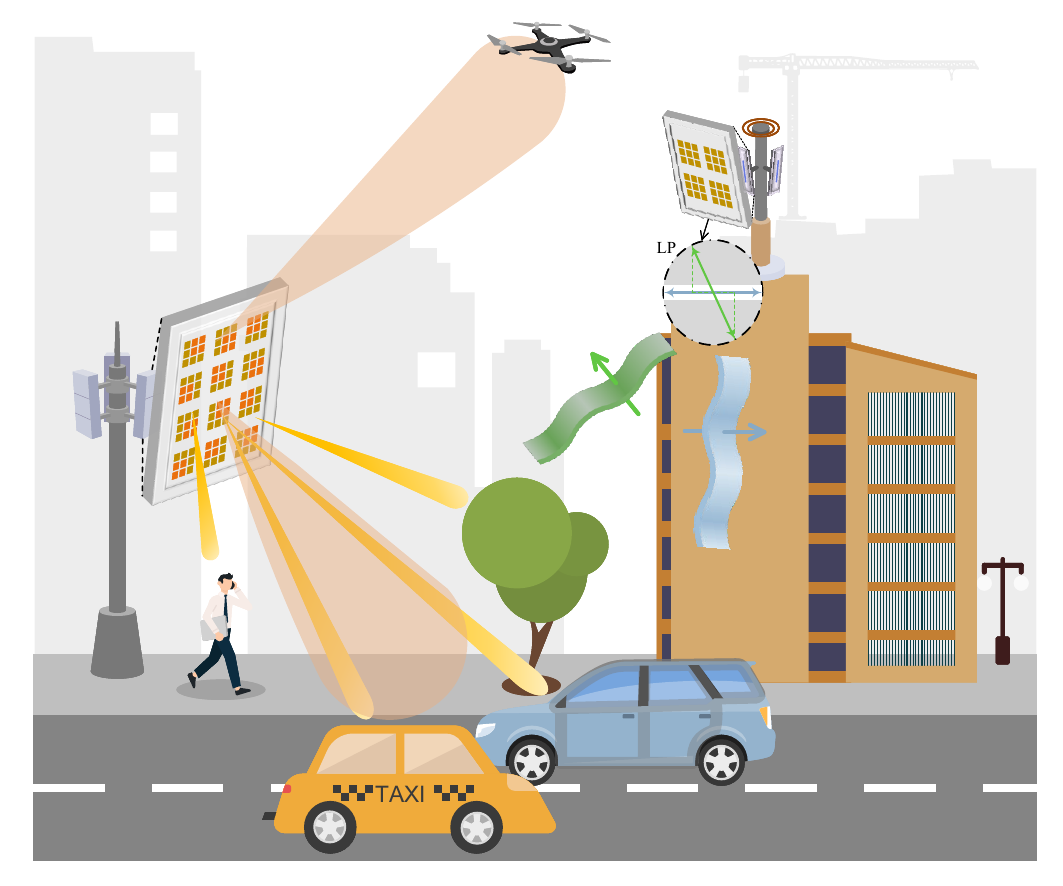}}
    \hspace{0.4cm}
    \vspace{-0.15cm}
    \subfigure[PLS.]{\includegraphics[width= 1.7 in]{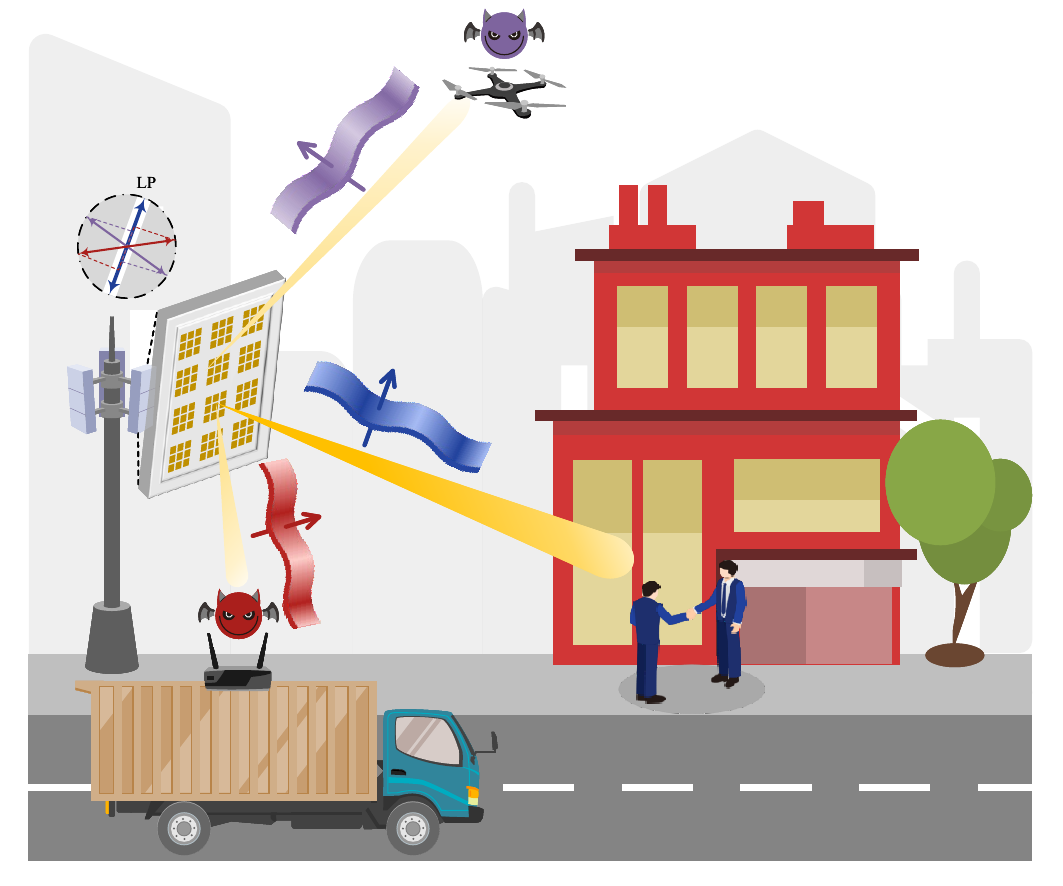}}
    \hspace{0.4cm}
    \vspace{-0.15cm}
    \subfigure[Near-field.]{\includegraphics[width= 1.7 in]{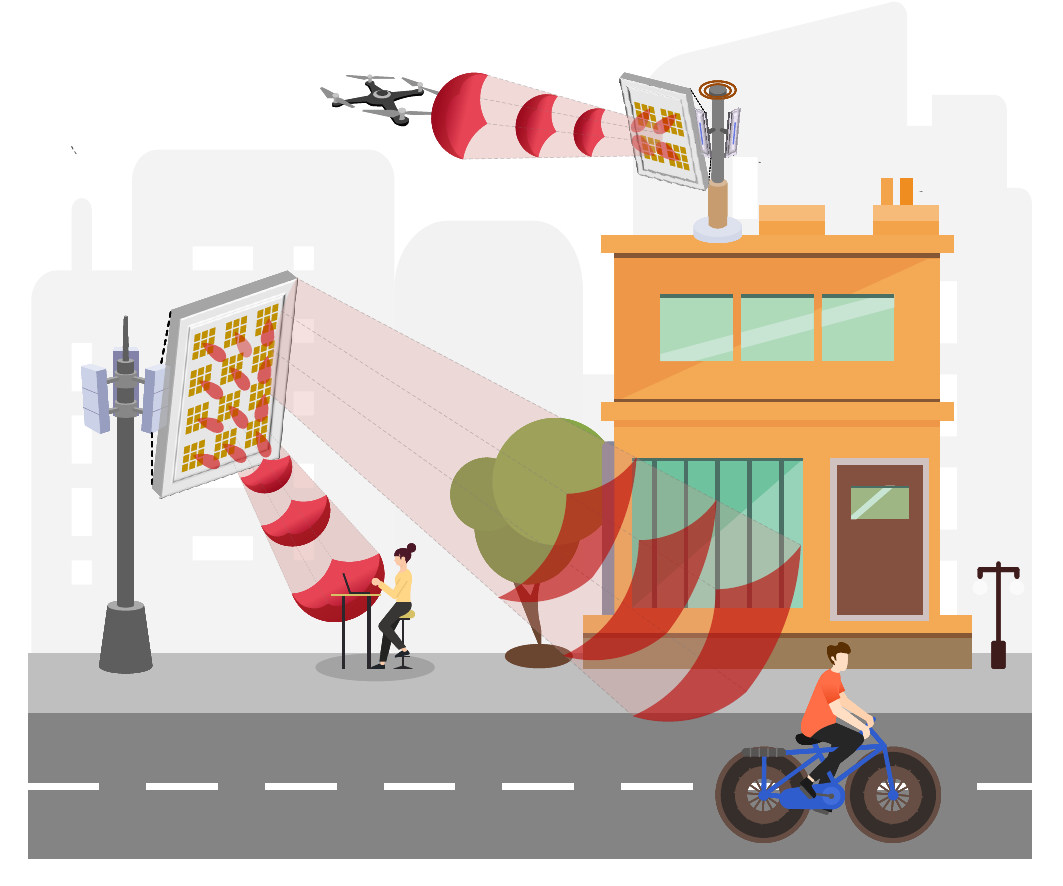}}
    \hspace{0.4cm}
    \vspace{-0.15cm}
    \subfigure[Full-duplex.]{\includegraphics[width= 1.7 in]{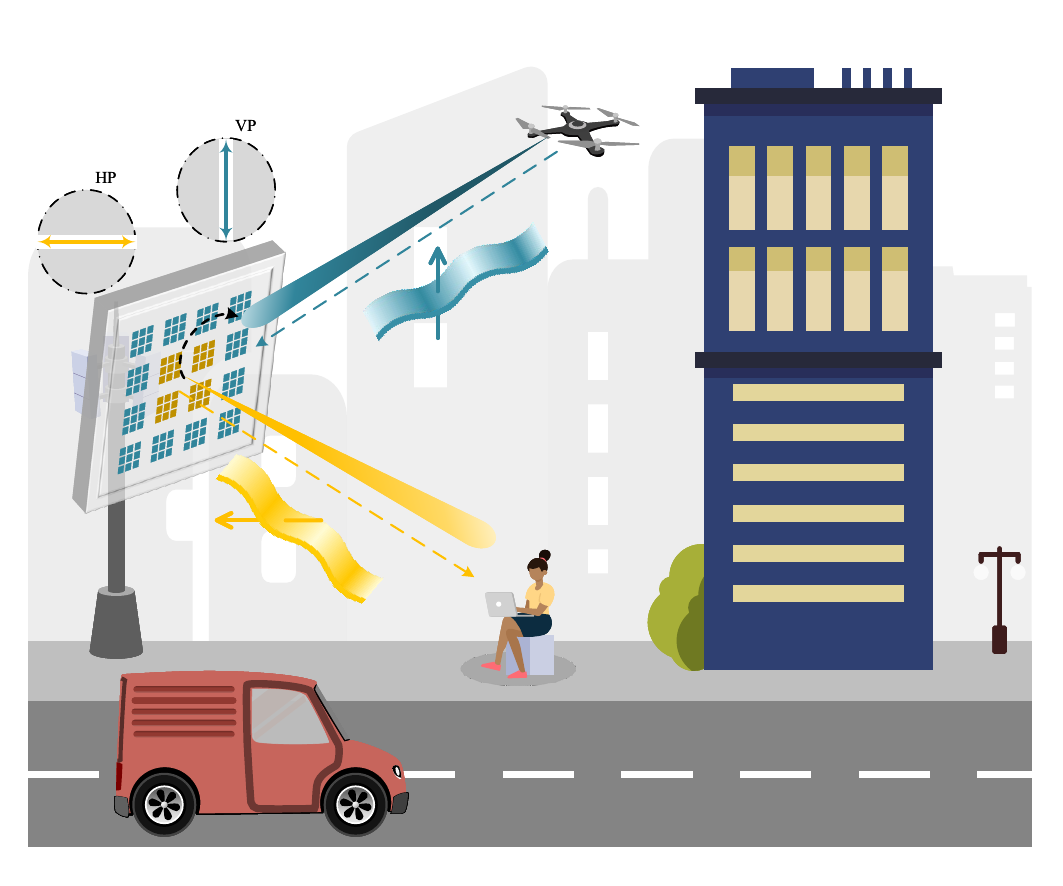}}
    \hspace{0.4cm}
    \subfigure[RA-based RIS.]{\includegraphics[width= 1.7 in]{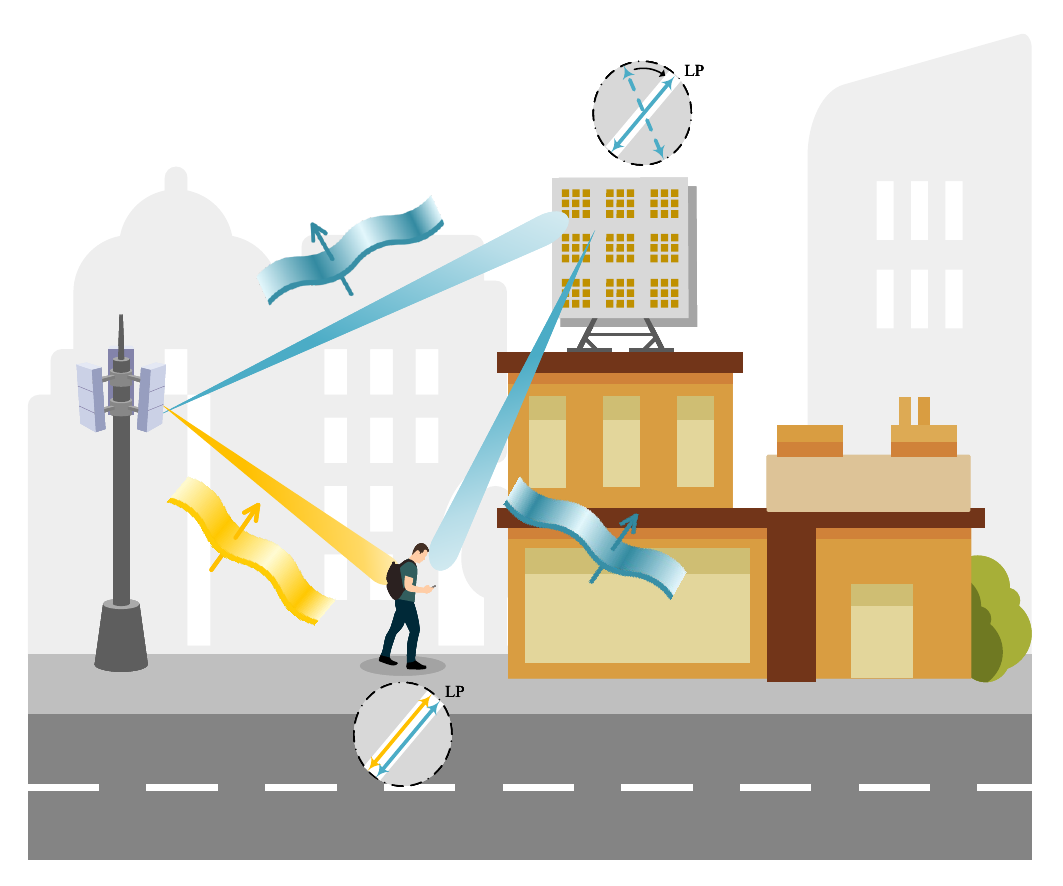}}
    \hspace{0.4cm}
    \subfigure[RA-based UAV/vehicular.]{\includegraphics[width= 1.7 in]{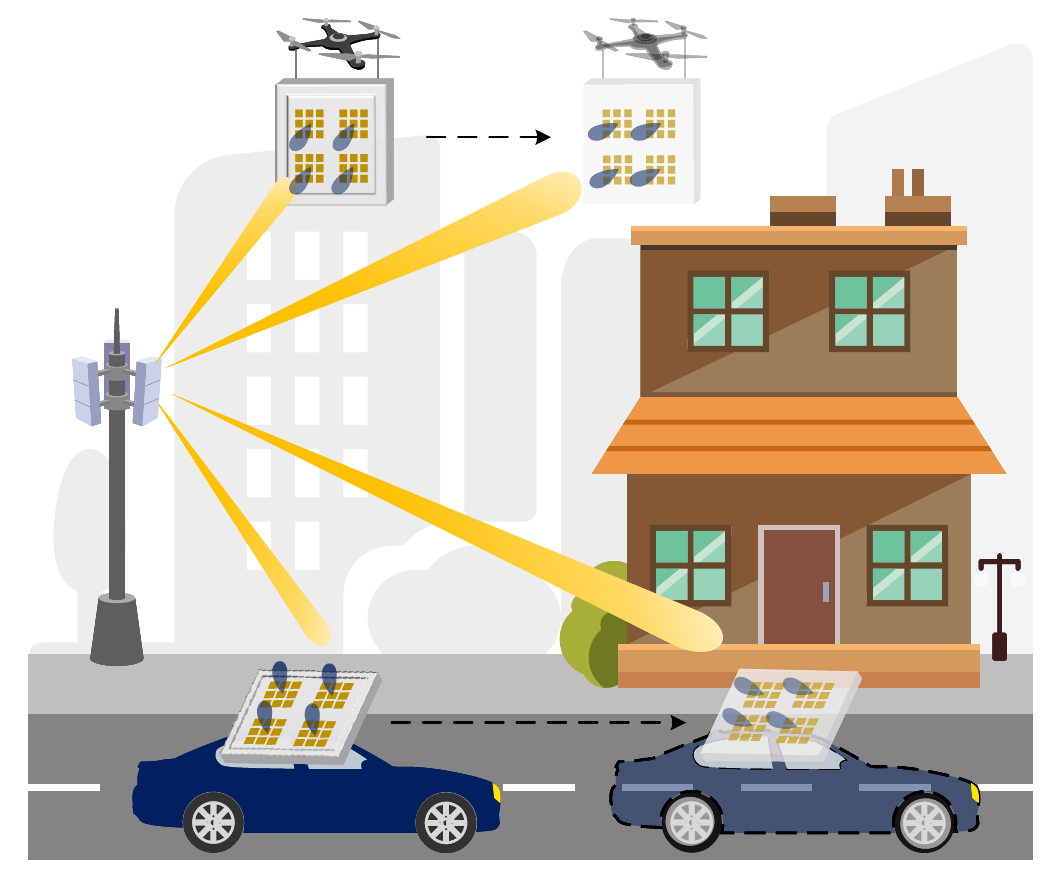}}
    \vspace{-0.1cm}
    \caption{Potential applications of RA arrays.}
    \label{fig:applications}
    \vspace{-0.5cm}
\end{figure*}

As wireless systems scale to higher frequencies and larger apertures, near-field effects become increasingly significant. In such regimes, the conventional plane-wave assumption breaks down, necessitating more accurate spherical wavefront models. This shift gives rise to spatial non-stationarity, wherein each antenna element experiences distinct channel characteristics, such as variations in radiation intensity and polarization orientation. These effects present challenges for conventional static antenna arrays, as the mismatch between fixed radiation patterns and spatially varying channel responses across the array can severely degrade energy focusing efficiency.
In contrast, RA arrays offer an effective solution by enabling per-element EM adaptability. By jointly optimizing EM radiation modes and SP strategies, RA arrays can shape wavefronts to match spatially varying environments, improving energy focusing and boosting performance in near-field scenarios, as depicted in Fig.~\ref{fig:applications}(c).

%As wireless systems move toward higher frequencies and employ larger antenna apertures, near-field effect becomes increasingly prominent. In these conditions, the classical plane-wave assumption no longer holds, necessitating the use of spherical wavefront models for accurate EM characterization. Consequently, spatial non-stationarity arises: Each antenna element experiences distinct channel properties, such as varying radiation intensity and polarization orientation. These effects pose critical challenges for conventional static antenna arrays, as the mismatch between fixed radiation characteristics and spatially varying channel responses across the array can lead to inefficient energy focusing and severe signal attenuation.

%In contrast, RAs provide a compelling approach to overcoming these challenges by dynamically adapting their radiation characteristics in response to local channel conditions. Through joint optimization of EM-domain radiation parameters and signal-processing strategies, RA arrays can proactively align radiation modes with spatially varying propagation environments, thereby achieving precise wavefront shaping and enhanced energy focusing capabilities. Such multi-domain coordinated design, as illustrated in Fig.~\ref{fig:applications}(c), not only mitigates spatial non-stationarity effects but also significantly improves system-level performance and efficiency in near-field communication scenarios.

\vspace{-0.4cm}
\subsection{Other RA Array-based Applications}
In addition to the scenarios discussed above, RA arrays support a variety of other promising applications in advanced wireless systems.
%RA arrays further support a range of emerging wireless applications:
\subsubsection{Full-duplex}
In full-duplex systems, integrating transmit and receive antennas into a compact single-array structure is desirable for reducing hardware complexity and conserving space. However, such integration inherently introduces severe self-interference that is inadequately addressed by conventional isolation or electromagnetic shielding methods alone. To effectively mitigate self-interference, the polarization reconfigurability of RA arrays can be employed, leveraging the orthogonality of transmit and receive polarization states, as illustrated in Fig.~\ref{fig:applications}(d).

\subsubsection{RA Array-based RIS}
Beyond their deployment in a base station (BS), RA arrays can also be effectively integrated as part of a reconfigurable intelligent surface (RIS). By employing polarization-reconfigurable elements, RIS can dynamically adjust the polarization state of the reflected signals, effectively reconstructing the polarization characteristics of the cascaded channel and significantly enhancing end-to-end link quality, as depicted in Fig.~\ref{fig:applications}(e).

\subsubsection{RA array-based UAVs/vehicles}
In unmanned aerial vehicle (UAV) or other vehicular communications, antenna arrays must be compact and lightweight due to stringent size and payload constraints. In addition, the high mobility of these platforms requires maintaining stable communication quality in rapidly changing channel conditions. RA arrays effectively meet these demands by offering miniaturized structures that dynamically adjust radiation patterns to proactively adapt to varying channel conditions, as shown in Fig.~\ref{fig:applications}(f). This capability maintains a quasi-stationary channel state, thereby ensuring stable reception and robust links in highly dynamic environments.

\vspace{-0.3cm}
\subsection{Case Study}
To demonstrate the effectiveness of RA characteristics in enhancing wireless system performance, this subsection presents simulation results for both communication and sensing scenarios involving RA arrays. We consider a BS equipped with $N = 8$ transmit/receive antennas that communicates with $K=2$ users and performs radar detection of a single target in the presence of $C=2$ strong signal-dependent clutter sources. All users and targets are assumed to be randomly located within an annular region centered at the BS, spanning angles betwen $[-\pi/3,\pi/3]$ and bounded by inner and outer radii of 30m and 60m, respectively.
For the communication scenario, user channels are generated by the Quasi Deterministic Radio Channel Generator (QuaDRiGa) under the ``3GPP\_3D\_UMi\_LOS'' setting, with
$L=5$ multipath components per user. For the sensing scenario, we adopt a classical line-of-sight (LoS) channel model that incorporates both path loss and depolarization effects. The path loss is modeled as $\alpha=10^{-{C_{0}}/{10}}\left({r}/{D_{0}}\right)^{-\kappa}$, where $r$ denotes the distance, $\kappa=2.2$ is the path-loss exponent, $C_{0}=30$dB, and $D_{0}=1$m. The depolarization (scattering) matrices are modeled as zero-mean complex Gaussian random matrices, with covariance structures consistent with those in \cite{M. Liu Security}. In the EM domain, the pattern-RA supports seven discrete radiation patterns \cite{R. Murch 2022 pixel fig_reference_round}, while the  polarization-RA offers four discrete polarization states \cite{H. Li 2024 polarization}. Other simulation parameters follow the settings in \cite{M. Liu Security}. We aim to jointly optimize the EM-domain radiation modes and the BB-domain digital beamformer to maximize either the communication sum-rate or the radar signal-to-clutter-plus-noise ratio (SCNR), subject to discrete EM mode selection and total transmit power constraints.

\begin{figure}[t]
    \centering
    \vspace{-0.0cm}
    \subfigure[]{\includegraphics[width= 3 in]{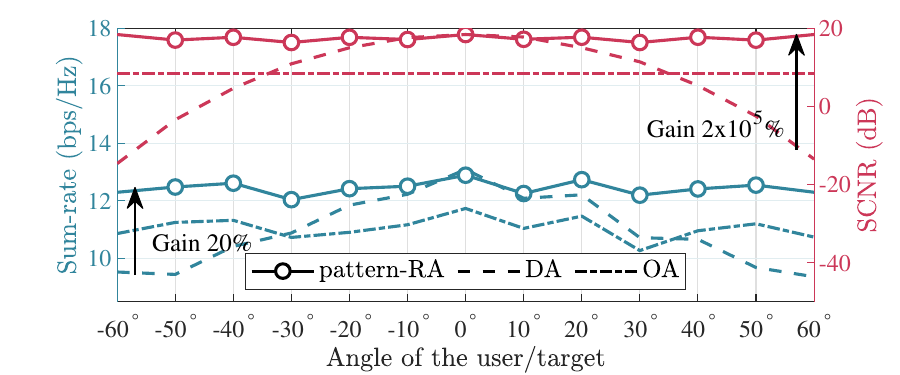}}%
     \vspace{-0.1cm}
    \subfigure[]{\includegraphics[width= 3 in]{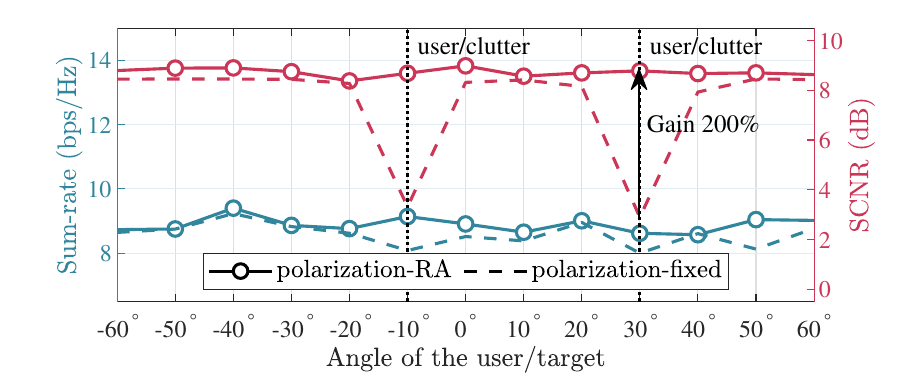}}
    \vspace{-0.4cm}
      \caption{Performance gains of pattern-RA and polarization-RA.}
     \vspace{-0.2cm}
    \label{fig:pattern_polarization}
\end{figure}

Fig.~\ref{fig:pattern_polarization} illustrates the impact of user/target angular position on communication sum-rate and radar SCNR for systems employing pattern-RA and polarization-RA, respectively. In Fig.~\ref{fig:pattern_polarization}(a), three array architectures are compared: pattern-RA, directional antennas (DA), and omnidirectional antennas (OA). In both communication and sensing scenarios, the pattern-RA-based scheme maintains consistently high performance by dynamically adapting its radiation pattern to the target/user angle. In contrast, the DA-based array performs well only near its principal beam direction, while the OA-based array offers uniform but relatively modest performance across all angles.
Notably, pattern-RA provides more significant gains in sensing than in communication. This is because sensing typically relies on a single line-of-sight (LoS) path, making it highly sensitive to angular mismatch, whereas communication benefits from multipath propagation, which offers angular diversity and mitigates this sensitivity. These results highlight the robustness and flexibility of pattern-RA-based architectures, demonstrating their effectiveness in reducing angle-dependent performance variability.

Fig.~\ref{fig:pattern_polarization}(b) further evaluates the role of polarization-RA. In this scenario, the angular position of one user/target varies, while the positions of the other users and clutter remain fixed. This setup illustrates the ability of polarization-RA to exploit polarization-dependent scattering characteristics for performance enhancement. In communication, the presence of multipath allows spatial separation even when users are angularly coincident, limiting the benefits of polarization diversity for user separation. However, in sensing with only an LoS path, if the target and clutter are angularly aligned, conventional polarization-fixed approaches that rely solely on BB-domain signal processing fail to separate them, resulting in significant performance degradation. In contrast, the polarization-RA-based approach can fully exploit differences in polarization to preserve detection performance under such challenging conditions.

\begin{figure}[t]
    \centering

   \includegraphics[width=3 in]{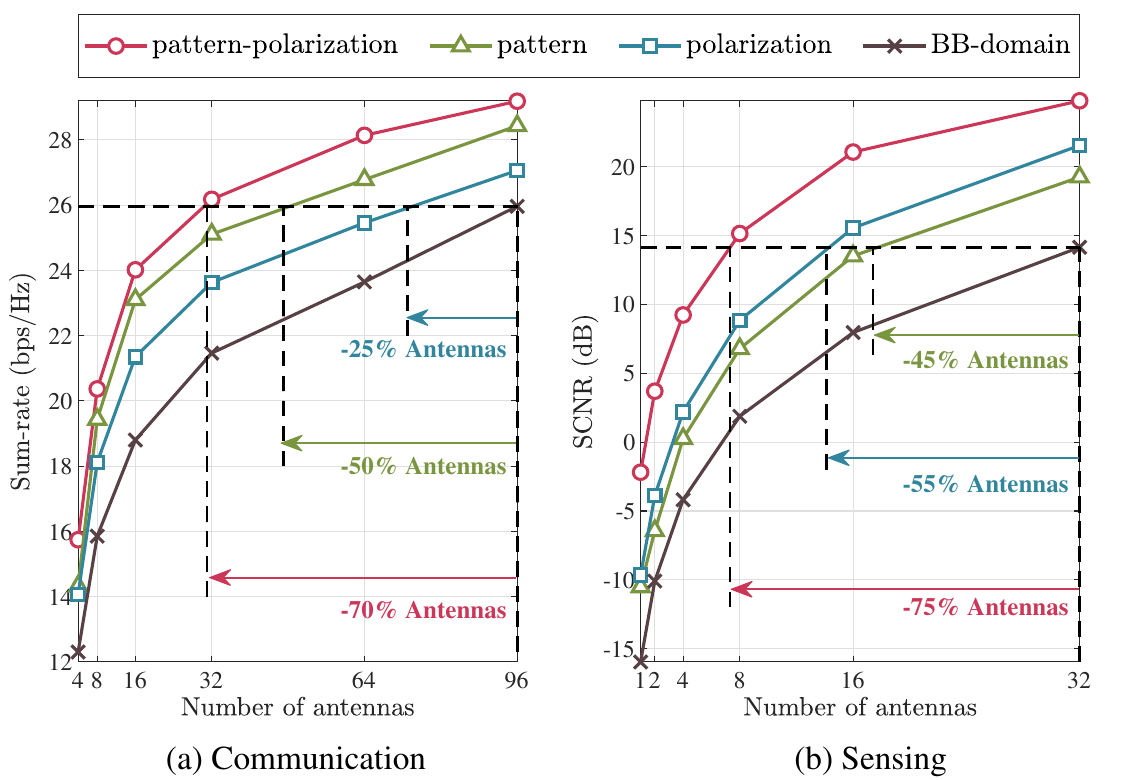}%
    \caption{Performance evaluation vs. the number of antennas.}
      \vspace{-0.5 cm}
    \label{fig:SCNR}
\end{figure}

Fig.~\ref{fig:SCNR} presents the performance evaluation as a function of the number of BS antennas in both communication and sensing scenarios, quantitatively demonstrating the inherent advantage of reconfigurability in terms of array size. In particular, the RA array with multi-domain coordinated reconfiguration achieves performance comparable to that of conventional antenna arrays employing BB-domain SP with up to 75\% fewer antennas. These results clearly highlight the hardware efficiency and size-reduction benefits of RA arrays, reinforcing their potential to support energy-efficient next-generation wireless systems without compromising overall performance. Moreover, the results suggest that pattern reconfigurability has a more pronounced impact in communication scenarios, whereas polarization reconfigurability plays a more critical role in sensing applications.

\vspace{-0.4cm}
\section{Open Challenges and Future Directions}\label{Future}
While RA arrays offer substantial potential to enhance wireless systems, several critical challenges must be addressed to enable their practical deployment. This section outlines key technical barriers and highlights promising research directions for advancing RA array-enabled wireless technologies.

\vspace{-0.2cm}
\subsection{Comprehensive Modeling, Theoretical Analysis, and Practical Measurement}

\subsubsection{Comprehensive Modeling}
Developing accurate and generalizable models that capture the diverse EM behaviors of RAs is a foundational requirement. Existing models are often overly simplified, scenario-dependent, or specific to particular reconfiguration modes, and thus fail to capture the intricate relationship between EM modes and SP across both circuit and system levels. Crucially, practical effects are often neglected, such as mutual coupling among array elements, component switching mechanisms, frequency-dependent radiation characteristics, and hardware-induced nonlinearities. To bridge this gap, comprehensive models that incorporate these phenomena are essential for reliable performance analysis, robust optimization, and practical deployment of RA-equipped systems.

%Developing accurate and unified models that comprehensively characterize RA array behaviors across diverse EM properties is crucial, especially for capturing the intricate coupling effects between EM modes and signal processing across both circuit and system levels. However, current modeling approaches often remain scenario-dependent, mode-specific, and excessively simplified, lacking the generality and depth necessary to accommodate various deployment scenarios. Specifically, many existing models neglect critical practical factors, including mutual coupling among antenna elements, detailed switching mechanisms within reconfigurable components, frequency-dependent radiation characteristics, and hardware-induced nonlinearities. Consequently, establishing holistic models that fully incorporate these factors is essential for achieving reliable performance evaluation, robust parameter optimization, and effective practical deployment of RA array-based wireless communication systems.

\subsubsection{Theoretical Performance Analyses}
Despite recent progress in RA design, theoretical frameworks for quantifying their impact on fundamental performance metrics remain insufficiently developed. In particular, deriving analytical bounds on capacity, outage probability, sensing accuracy, and estimation error in the presence of EM-domain DoFs remains an open research challenge. Furthermore, identifying the critical operating regimes where RA array-enabled systems outperform  their conventional counterparts can reveal deeper insights into the unique performance advantages afforded by RA technologies.

%Despite increasing attention towards RA technology, comprehensive theoretical analyses that rigorously quantify its impact on fundamental performance metrics remain limited. In particular, deriving explicit upper and lower bounds for performance metrics, such as channel capacity, outage probability, and sensing resolution in the presence of additional EM-domain DoFs remains an open research area. Furthermore, exploring and characterizing the performance trade-off regions associated with RA-equipped versus conventional antenna configurations is essential to accurately evaluate and demonstrate the inherent potential offered by RA arrays.

\subsubsection{Practical Deployment and Measurement}
Although simulations provide valuable preliminary insights, empirical validation through real-world experimentation is vital for demonstrating practical feasibility. Experimental platforms that allow real-time dynamic radiation pattern control during data transmission and reception are needed. Additionally, systematic measurement campaigns across diverse propagation environments and hardware setups are required to assess performance consistency, robustness to hardware impairments, and reconfiguration latency. Ultimately, rigorous prototyping and field trials are indispensable for translating RA innovations into practical system gains.

%While theoretical analyses and simulations yield valuable insights into RA capabilities, practical validation and experimental verification are still substantially lacking. To bridge the gap between theoretical predictions and practical implementations, it is essential to develop experimental platforms that enable real-time dynamic control of antenna radiation modes during active data transmission and reception. Additionally, extensive measurement campaigns covering a variety of propagation environments and hardware setups are necessary to evaluate critical system-level factors, such as performance consistency, robustness to hardware imperfections, and switching latency. Ultimately, rigorous prototyping and systematic field trials are indispensable for assessing the practical feasibility and tangible performance improvements achievable with RA technology in realistic wireless scenarios.

 \vspace{-0.2cm}
\subsection{Hardware Non-idealities and Coupling Effects}

\subsubsection{Intrinsic Component Nonlinearities}
RA arrays rely on hardware components such as PIN diodes, varactors, and FETs, which exhibit non-ideal behaviors including hysteresis, thermal drift, and saturation effects. These nonlinearities can distort radiation properties, degrade impedance matching, and cause unintentional frequency shifts, thereby undermining intended EM responses. Coordinated cross-domain design under such imperfections may necessitate advanced calibration techniques, pre-compensation methods, and adaptive control mechanisms to maintain desired performance.

\subsubsection{Inter-RA Mutual Coupling}
While coupling between active and parasitic elements is deliberately designed into a single RA to generate desired radiation modes, unintended mutual coupling among RA elements poses significant challenges, especially in compact or dense arrays. Such detrimental coupling can distort radiation modes, impair beamforming precision, and compromise system reliability. As emphasized in Section V-A, accurate EM-domain modeling of this effect is critical. Compensation algorithms that jointly optimize antenna configuration and signal processing must be developed to maintain robust operation under coupling-induced distortions.

\vspace{-0.2cm}
\subsection{Channel Estimation}

\subsubsection{Impact on Conventional Estimation}
RA-induced variations in radiation characteristics introduce non-stationarity in the channel, thereby complicating the acquisition of accurate channel state information (CSI). The coupling between EM parameters and spatial channel features makes traditional estimation approaches inefficient, as exhaustive measurements for all radiation modes are infeasible. Recent advances in channel extrapolation leverage inter-mode correlations to infer unmeasured CSI, and tensor decomposition techniques to exploit the high-dimensional structure of coupled channels and  enhance estimation performance with reduced pilot overhead.

\subsubsection{Radiation-Pattern-Aware Codebook Design}
As previously discussed, RA arrays induce varying radiation characteristics across antenna elements in an array that disrupt the orthogonality of conventional codebooks, leading to degraded beam training accuracy. Traditional beam
codewords may become mismatched in RA arrays, impairing received signal strength assessment and beam alignment. Thus, it is essential to develop radiation-pattern-aware codebooks where each codeword jointly encodes a beamforming direction along with specific antenna mode configurations. Such designs enable orthogonal codebook structures tailored to RA features, thereby supporting accurate and low-overhead beam training.

\subsection{Cross-Domain Designs}

\subsubsection{Optimization-based Algorithms}
The joint optimization of EM configurations and SP in RA array-enabled systems yields complex formulations characterized by high dimensionality and tightly coupled variables. Modeling the intricate interactions between the EM and signal domains adds complexity and conventional decomposition techniques are often inadequate for such problems. Thus, efficient and scalable optimization methods are thus essential for practical real-time operation.

\subsubsection{AI-driven Algorithms}
The large control space and dynamic nature of RA configurations render classical optimization approaches computationally burdensome. Moreover, the relatively slow hardware reconfiguration speeds demand predictive methods to enable timely adaptation. Artificial intelligence (AI)-based solutions, including supervised deep learning and reinforcement learning, offer effective alternatives for low-complexity computation and RA mode prediction. Future efforts should focus on developing lightweight, hardware-friendly AI models that support fast inference and robust adaptation in dynamic environments.

\vspace{-0.2cm}
\subsection{Cell-free Networks with RA arrays}
Cell-free networks rely on densely distributed access points (APs) to ensure uniformly high spectral efficiency across the coverage area. However, dense deployments demand compact, energy-efficient hardware capable of wide-angle beamforming, which poses challenges for conventional large-array solutions. RA arrays provide an effective alternative by enabling compact, low-power APs that retain strong beamforming capabilities via EM reconfigurability. This allows RA-equipped APs to emulate massive MIMO performance with fewer elements, offering a scalable and cost-efficient solution for practical cell-free deployments.

Despite these advantages, integrating RA arrays into cell-free networks introduces new algorithmic challenges. Coordinating RA-equipped APs requires tailored optimization frameworks that account for their reconfigurable and multi-domain nature. Centralized approaches can achieve globally optimal RA mode selection and joint signal processing but suffer from scalability limitations and high communication overhead. In contrast, distributed algorithms offer better scalability and lower overhead but must contend with partial channel state information (CSI) and limited inter-AP coordination. Future research should focus on hybrid algorithms that combine the strengths of both centralized and distributed designs, enabling efficient and scalable coordination of RA-equipped APs.

\section{Conclusions}
In this article, we have investigated the transformative potential of RA arrays as a foundational technology for next-generation wireless systems. We first reviewed the  operating principles and radiation characteristics of individual RA elements. Building on this foundation, we introduced advanced RA array architectures that enable coordinated multi-domain beamforming. A set of representative applications, along with a detailed case study, demonstrated the flexibility and practical advantages of RA array-enabled systems. Finally, we outlined several open research challenges and future directions that are essential to fully realize the capabilities of RA arrays in intelligent, adaptive, and high-capacity wireless communication environments.

\section*{Biographies}
\vspace{-1 cm}
\begin{IEEEbiographynophoto}
{Mengzhen Liu} (liumengzhen@mail.dlut.edu.cn) is now studying toward the Ph.D. degree with the School of Information and Communication Engineering, Dalian University of Technology.
\end{IEEEbiographynophoto}
\vspace{-1 cm}
\begin{IEEEbiographynophoto}
{Ming Li} (mli@dlut.edu.cn) is presently a Professor at the School of Information and Communication Engineering, Dalian University of Technology.
\end{IEEEbiographynophoto}
\vspace{-1cm}
\begin{IEEEbiographynophoto}
{Rang Liu} (rangl2@uci.edu) is currently working as a Postdoctoral Scholar  at the Department of Electrical Engineering and Computer Science, University of California, Irvine.
\end{IEEEbiographynophoto}
\vspace{-1 cm}
\begin{IEEEbiographynophoto}
{Qian Liu} (qianliu@dlut.edu.cn)  is currently a Professor at the School of Computer Science and Technology, Dalian University of Technology.
\end{IEEEbiographynophoto}
\vspace{-1 cm}
\begin{IEEEbiographynophoto}
{A. Lee Swindlehurst} (swindle@uci.edu)  is currently a Professor  at the Department of Electrical Engineering and Computer Science, University of California, Irvine.
\end{IEEEbiographynophoto}

\end{document}